# Social Network Analysis Taxonomy Based on Graph Representation

Andry Alamsyah[1], Budi Rahardjo[2], Kuspriyanto[2]

[1] School of Electrical Engineering and Informatics, Institut Teknologi Bandung
School of Management Technology and Media, Institut Manajemen Telkom
andrya@telkomuniversity.ac.id
[2] School of Electrical Engineering and Informatics, Institut Teknologi Bandung
br@paume.itb.ac.id; kuspriyanto@yahoo.com

*Abstract*

*There are three approaches in current social network analysis study: Graph Representation, Content Mining, and Semantic Analysis. Graph Representation has been used for analyzing social network topology, structural modeling, tie-strength, community detection, group cohesion visualization, and metrics computations. This paper provides a taxonomy of social network analysis based on its graph representation.*

## I. Complex Network and Social Network Analysis

In the last few years, we experience unprecedented growth usage of online social media. Together with the popularity of web 2.0, they support the development of *Social Computing* [37], which is an area in information technology to study human behavior and social relations connected via computer networks. The demand of exploration and exploitation social interactions in online social media is very high, that triggered a new research field. Social computing is also defined as intersection between computer science and social science [37]. The early research of *Social Network Analysis* (SNA) started in the 60's when it was difficult to do the large-scale data experimentation because of the limitation of computing power. Today by using online social network, we have massive amount of data that makes us possible to reveal social structures, to model social relations, to facilitate information exchange between individual or between users inside the group. SNA draws a lot of attention recently due to the ability to quantify social networks, it affect human understands better their social and network and its implications. Social media is also widely used as a platform for information dissemination and many research in computational science, management, sociology, and many others areas put their effort to understand how its work.

The majority of research in SNA is using *Graph Theory* [28][33]. The Traditional *Sociogram* [33] brings the first idea to use graphic representations of social links, where nodes represent actors or entities while edges represent relations between actors or entities. Social structures is built from social links contains individual or organizational connected one and another by means of friendship, kinship, interest, financial transaction, like/dislike, trust, beliefs, sexual relationship, knowledge, prestige and many others. Those relationship can be viewed as a graph representation either symmetric or asymmetric relations in a form of "ties", "links", "connections" called *Network Theory* [10]. There are many applications of Network Theory in many disciplines such as computer science, biology, management, economy, statistical physics, particle physics, operations research and sociology. The graph representation model of social network is used for exploring network features, most influential actors in social network, structures and network topology. In Fig 1, we give an example of social network visualization based on friendship between 34 members of a karate club in graph representation using data set from Zachary [43].

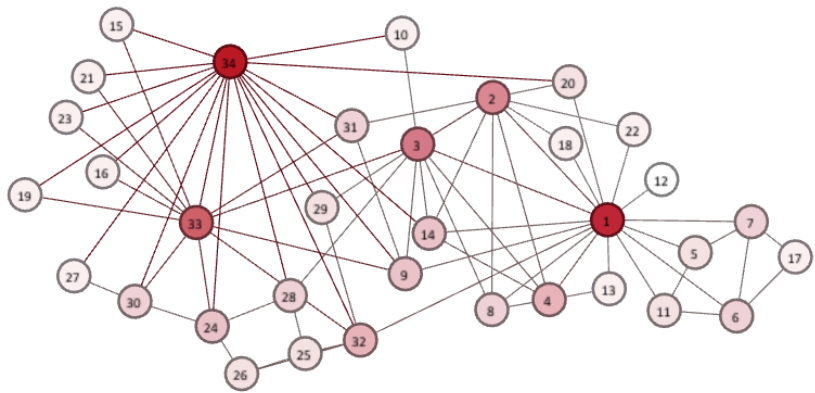

*Fig 1. Graph representation of social networking between 34 members of karate club, data set collected by Zachary [43]*

The real world network formation is not as straightforward as lattice graph or random graph but it is more likely formed a *Complex Network* [2] [7] [10], which has non-trivial topological features, heterogeneity, interactions among essential nodes of the systems complicate dynamical rules and without a global supervision. The behavior of real networks very distinct from traditional assumption, which is supposed to have majority of nodes about the same number of connections around an average. This is a typically modeled by random graphs. The modern network research shows the majority of nodes of real networks is very low connected, and there exist some nodes with very high connectivity / hubs. This is a distinct feature of complex networks that we called *power law* or *scale free* characteristics [2] [7] [10]. Some examples of complex networks are technological / engineered networks such as telecommunication networks [31], internet, power grids, transportation networks, delivery and distribution networks; information networks such as world wide web, citation networks, peer to peer networks, recommender networks [39]; biological networks [19] such as biochemical networks, neural networks, ecological networks; social networks such as affiliation networks, cognitive and semantics networks, small world features, economics and market behavior [22].

The study of complex network is called *Network Science* [28], which is an interdisciplinary academic study from mathematics, statistical mechanics, inferential modeling, social structures, data mining, and information visualization [29], with the purpose for prediction model for each phenomenon in any type of the network. New interest and research on network science particularly focus on network whose structures is irregular, complex, dynamically evolving over time, with the main focus the analysis of small network to that system with thousands or millions nodes [2]. *Network Properties* [29] is the main underlying idea to model our network. Some of classical network properties are *density, size, average degree, average path length, diameter, clustering coefficient, components, core, cliques, connectedness, centrality* and many more complex properties [20] such as *maximum flows, Hubbell / Katz cohesion.* To date there are several *Network Models* based on the complexity, function and timeline social network development, some of them are: *Erdos-Renyi* [7] model for generating random network models, *Barabasi-Albert* [10] model for growing real world network based on two assumptions growth and preferential attachment, *Watts-Strogatz* [42] model for random network model with small-world properties. Understanding many aspect of complex network may serve our knowledge to SNA concept comprehensively.

## II. Social Network Analysis Taxonomy

Taxonomy is important to acknowledge the latest boundary of SNA study, mapping the different on-going efforts and research interest in SNA topics. Our literature review include adaptation of the the classical SNA, identifies popular models used by researchers for representing and visualizing social networks, analyze current and future SNA development. We see that SNA-based research have tendency give bigger role in development of content analysis and semantic models. The needs of extracting meaning / context on online social network drives the effort of finding semantics effect in social network research. There are three major approach of SNA study [9]: *Graph Representation*, *Content Mining*, *and Semantic Analysis*. Graph representation is the SNA-graph based from the classical approach such as sociogram to the latest research development such as *community detection, network structures, random walks and temporal networks*. Content Mining focuses on understanding the models and identifying factors that drive information dissemination in online social network. Several factors to be consider such as hashtag, URL, sentiment analysis, sarcasm detection. Semantic analysis in SNA is developed due to lacking semantics support in graph representation and content analysis. Recent development of semantics SNA is incorporation semantics in rich structured data, increasing semantics awareness capturing social networks in much richer structures than raw graph [14]. However in this paper we focus our SNA taxonomy based on graph representation due to some opportunity available, especially in handling large-scale data and temporal networks. Our illustration of the SNA taxonomy based on graph representation is shown in Fig 2.

We categorize SNA based graph representation into 5 major areas according different approach on issues that they are addressing. They are *Metric, Network Structure, Random Walks, Temporal Graph and Visualization.* Metrics concern with network measurement, network properties and network quantification. Network Structure focuses its study on network topology and its features. Random Walks study the walk-through network by following path at random, this study is to understand how information is spreading across the network. Temporal Networks focuses its study on how to deal with network that its nodes are not always active, which is very natural in real world. Visualization is important to understand network data and convey the analysis result. It is the most natural model that people want to look at their network and often used as an additional or standalone data analysis methods. The details of each area will be presented in the following chapter.

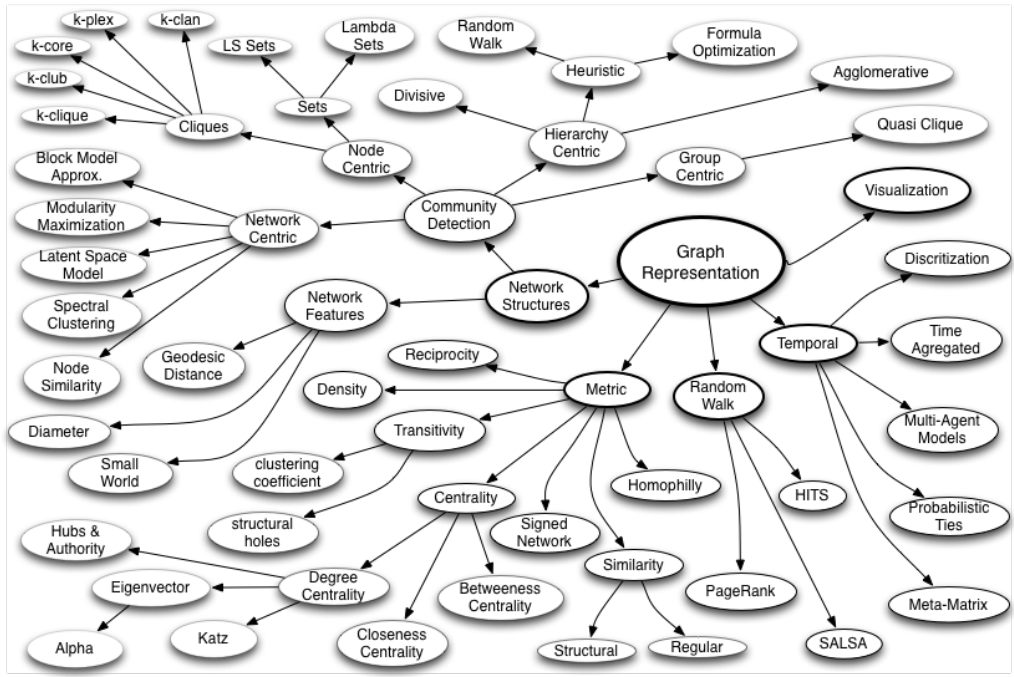

*Fig 2.*
*Social network analysis taxonomy based on graph representation*

III. Social Network Analysis Metrics

SNA Metrics is one of the most extensive researches in social network and graph representation area. These metrics based on the classical sociogram and borrowed from mathematical graph formula. Some network properties we mention previously also actively contribute in forming metrics criteria. By knowing relations between nodes inside a network, we can calculate varieties useful measures and importantly metrics can reveal network structures. Having metrics also means we quantify a network and use the measurement for several purposes. Essentially we divide metrics based on the popularity of SNA application in real world, they are *centrality* and *non-centrality*. Centrality is widely used in many applications and it measures the degree to which network structures determine the importance of a node in a network [24]. We divide centrality into three parts: *degree centrality, betweenness centrality* and *closeness centrality*. Non–centrality metrics are including *reciprocity, density, transitivity, homophilly, component, similarity,* and *signed networks*. Degree centrality basic idea is to compute how many edges are tied to a node, with variations on how we distribute score proportionally to their neighbors, how we compute on directed / undirected network, how we rank the importance of node based on in / out-degree. These variations determine the following metrics

*Eigenvector* (and its adaptation called *Alpha*)*, Katz, Page Ranks and Hubs & Authority*. The complexity of real world networks implicate measuring using degree centrality itself is not sufficient. To complete centrality measures in different aspect we may use betweeness and closeness. The details of centrality-based metrics can be seen in Table 1.

*Table 1. Centrality-based Metrics*

| Name | Description | Formulas |
|---|---|---|
| **Eigenvector** | Each node has proportional value to the sum of the score its neighbors. The value can be large because either a node has many neighbors and/or it has important neighbors. First proposed by Bonacich [3] | $x_i = \lambda^{-1} \sum_j A_{ij} x_j$ <br> where $x_i$ is the score at node i, $A_{ij}$ is the correspondent value on matrix adjacency, $\lambda$ is the eigenvalue. |
| **Katz** | Each node has given small amount of centrality "for free" regardless of its position in the network. It's a solution for eigenvector centrality on directed network. First proposed by Katz [28] | $x_i = \alpha \sum_j A_{ij} x_j + \beta$ <br> where $x_i$ is the score at node i, $A_{ij}$ is the correspondent value on matrix adjacency, $\alpha$ and $\beta$ are positive constant. |
| **PageRank** | Each node has given the rank based on network neighbors proportional to their centrality divided by their out-degree. First proposed by Page and Brin [4] | $x_i = \alpha \sum_j A_{ij} \frac{x_j}{k_j^{out}} + \beta$ <br> where $x_i$ is the score at node i, $A_{ij}$ is the correspondent value on matrix adjacency, $\alpha$ and $\beta$ are positive constant, $k'_{out}$ is the out-degree |
| **Hubs & Authority** | Authority is a node that contain useful information on topic of interest, while Hub is a node that tells us where the best authority to be found. To compute Hubs and Authorities we use HITS algorithm. | $x_i = \alpha \sum_j A_{ij} y_j$ , $y_i = \beta \sum_j A_{ji} x_j$ <br> where $x_i$ is the authority value and $y_i$ is the hub value |
| **Closeness Centrality** | Measure the mean distance from a node to other nodes by geodesic path. | $C_i = \frac{1}{n-1} \sum_{j(\neq i)} \frac{1}{d_{ij}}$ <br> where $C_i$ is the closeness centrality value at node i, $d_{ij}$ is the distance between node i and j, n is number of node in the network. |
| **Betweenness Centrality** | Measures the number of shortest paths going through a node. First proposed by Freeman [15] | $C_i = \sum_{i \neq s \neq t} \frac{\sigma_{st}(i)}{\sigma_{st}}$ <br> where $C_i$ is the betweenness centrality value at node i, $\sigma_{st}(i)$ is the number of shortest path between node s and t that pass through node i. $\sigma_{st}$ the number of shortest path between node s and t |

In non-centrality metrics there are wider approach on how to measure a network. The details of each metric can be seen in Table 2. The metrics are measuring connections between nodes, nodes distributions and group / component segmentations in a network

*Table 2. Non-Centrality-based Metrics*

| Name | Description |
| --- | --- |
| **Reciprocity** | In directed networks, the tendency of two nodes form mutual connections between each others |
| **Density** | The fraction of number edges in network to the maximum edges possible [33] |
| **Transitivity** | A likelihood that two associates of nodes are associates. *Clustering Coefficient* is a measure of a proportion number of pairs in a network that are connected to number of available pairs in the network. If the expected ties between neighbors missing. The missing links called *Structural Holes* and its first studied by Burt [5] |
| **Similarity** | A measure for comparing similarity between two/more networks. Similarity can be determined in many different ways; two most common are *structural equivalence* and *regular equivalence*. |
| **Component** | A measure of maximal subset of nodes such that each node is reachable by some paths from each others. |
| **Signed Networks** | A network in which their edges have signed either + or -, for example friendship have positive edge while animosity have negative edge. Networks containing only loops with even numbers of minus signs are said to show structural balance, which proved by Harary [17] |
| **Homophilly** | The tendency of nodes to form ties with similar nodes rather than dissimilar nodes. Ties can be friendships, acquaintances, business relations and others. Similarity can be gender, race, age, occupation, education, status and other characteristics. Homophilly also called as *Assortative Mixing* |

## IV. Network Structure

The objective of studying the network structures is to have better understanding on how are the structure formed and the mechanism that affects information diffusion, contagion, community finding and identification *cohesive subgroup* [40]. We divide network structures into *Network Features* and *Community Detection*, both are focus on how we explore network from its structures. In network features, we discuss some properties as follows: *Geodesic Distance* is generalization from *Geodesic Path* that also called simply as a shortest path, it is a path between two nodes such that no shorter path exists [25]. *Diameter* of the network is the length of the longest geodesic path between any pair of nodes in the network for which path actually exist [1]. Today, social network produce large-scale data that arise computational complexity, for example to find a geodesic path in a network with *n* nodes and *m* edges, the complexity is $O(m) + O(n)$ [26], the complexity to measure betweeness centrality is $O(n^3)$ [21] [27]. A feature like *small world phenomenon* [12] can reduce our complexity if we know how to find the shortcut between two separate nodes.

Community is created out of *bipartite network* where peoples meet at the same event, creating networks where we have multi modal nodes contains peoples and events. This will lead to *affiliation networks* where each actor have more than one group or in other word, they become member of more than one group, creating large complex community structures. *Hypergraph* can explain this phenomenon in graph theory. To detect a community [35][36], we use several scenarios as follows: *Mutuality of ties*, which means everybody in the groups, knows each other. This property is called *cliques*. *Frequency of ties* among members means everybody in the group has links to at least *k* others in the group. The properties of this measure are *k-core* and *k-plex*. *Closeness* or *reachability* to other member means individuals are separated by at most *k* hops. The properties for this measure are *k-clique, k-clan,* and *k-club*. And the last, *Relative frequency of ties* among group members compared to outside group members, describing nodes connected to at least proportion *p* of outside group members. The property is called *p-cliques*, and there are other properties for this scenario such as *LS sets* and *Lambda sets*.

The criteria for community detection may vary [36], but in general there are four methods categorization: *node-centric, group-centric, network-centric* and *hierarchy-centric*. In *node-centric*, each node must comply certain properties such as complete mutuality and reachability. In *group-centric* a group has to satisfy certain properties without look the details in every node. It is acceptable that some nodes in the group have low connectivity as long as in overall the group satisfies the properties. In *network-centric*, the goal of the methods is to create disjoint sets by partitioning network. Typically, network-centric community detection aims to optimize a criterion defined over network partition rather than over one group. *Node Similarity* see nodes are structurally equivalent if they connect to the same set of nodes, *Latent Space Models* transform nodes in simple dimension such as Euclidean Space to simplify the calculations, *Block Model Approximation* minimize the difference between interaction matrix and a block structure, *Cut Minimization* minimize the cut which is the number of edges that belong to outside groups and *Modularity Maximization* measure group interactions compared to the expected random connections in the group. In *hierarchy-centric,* the goal is to build hierarchical structure of community based on network topology. There two representatives approaches, those are *Divisive Clustering* that iteratively partitions nodes into smaller and smaller subset [16], and *Agglomerative Clustering* that initializes nodes to form communities and iteratively merges communities satisfying certain criteria into larger and larger communities [11]. There are others algorithm-based *Heuristics* such as random walks, analogies to electrical networks or formula optimization [16].

## V. Random Walks, Temporal Networks and Visualization

Random Walks is a path across a network created by taking repeated random steps. Starting at some specified initial node, at each step of the walk we choose uniformly at random between the edges attached to current node, move along the chosen edge to the vertex at its other end, and repeat. SNA approach using Random Walks can be found on snowball sampling which is the sampling for hidden populations [38]. Random Walks is also tightly connected with the role ranking algorithm, an efficient ranking algorithm is important in any retrieval system. With the huge number of website exist today, the availibility of search result rank based on user contexts is very crucial [32][34]. There are three ranking algorithms *PageRank* [4], *HITS* [22], and *SALSA* [23]. The propagation of *Trust / Influence* [39] in a network with alleviate cold start problem also follow the random walks based algorithm.

Most of the graph representation we study is static networks due to the non-trivial solution over *Temporal Network*, a network which the edges are not continously active [5]. Like the static network topology, the temporal structure of edge activations can affect dynamics of systems interacting through network, from disease contagion to information diffusion over an e-mail network. The study of temporal networks uses the framework from static network and analyze their inter-relation that affect the behavior of dynamical systems. There are several approches to measures temporal-topological structures [8], represent temporal data as a static graph, and model temporal networks : Graph *Discretization, Time Aggregated, Metamatrix, Probabilistic Tie*s and *Multi-agent Models.*

One of the most and basic features peoples need for network modeling is visualizing their network. *Visualization* is practical if we work on limited number of nodes and impractical as soon as our network become larger. Today there are many softwares that can help us visualize our network and calculate most metrics available, from limited number node to the scalable network. With respect to visualization, network analysis tools are used to change the layout, colors, size, and other network representation. We can see the current comprehensive list of SNA Software in [30]

## VI. Conclusion

This paper focusing on SNA taxonomy based on graph representation. We have examined overall social networks analysis, explain the formal methods available, presenting social network properties and mapping research categories. Categorization in SNA based graph representation is based on different approach on issues that they are addressing. In centrality-metrics, we have given comparison on each metrics, however their performance still need to be tested using real data. In non centrality-metrics, they are not comparable since each metric use different approach to solve the problem. In community detection, the absence of ground truth information about a community structure in real world network give rises many new methods other than we mention in this paper.

Eventhough SNA based on content mining and semantics analysis research-based catch a lot of attention lately for promising rich social network analysis, the researches based on graph representation are still promising. The adoption SNA approach into many real world application open new perspective on how modeling graph representation. Online social network such as facebook and twitter are connecting hundred millions of users, they create large-scale network structure available. This pose a challange on scalability, heterogeneity, evolution, collective intelligence, evaluation. Other challange is utilizing temporal networks approach for real world problem that can handle large multi-mode, multi-link networks with varying level of uncertainties.